# Magnetic and Transport Properties of Ternary Indides of type $R_2CoIn_8$ ($R$ = Ce, Pr and Dy)

Devang A. Joshi[1], C.V. Tomy[1], P.L. Paulose[2], R. Nagarajan[2],

R. Nirmala[2] and S.K. Malik[2]

[1]Department of physics, IIT Bombay, Mumbai 400 076, India

[2] Tata Institute of Fundamental Research, Mumbai 400 005, India

**Abstract**

We have synthesized and investigated the magnetic and transport properties of a series of compounds, $R_2CoIn_8$ ($R$ = rare earth). Compounds form in single phase with a tetragonal structure (space group $P4/mmm$, no. 162). The Ce compound shows heavy fermion behavior. The magnetic susceptibility of $Pr_2CoIn_8$ shows a marked deviation from the Curie-Weiss behavior at low temperatures, which is attributed to the crystalline electric field effects. Heat capacity and magnetization measurements show that $Dy_2CoIn_8$ undergoes a magnetic transition at 17 K and a second transition near 5 K, the latter of which may be due to spin reorientation. Magnetization of this compound shows two metamagnetic transitions approximately at 3.6 T and 8.3 T.

Keywords: Heavy Fermion, Crystalline electric field effects, metamagnetic transitions

## 1. Introduction

The ternary indides of the general formula, $R_nMIn_{3n+2}$ ($R$ = rare earth; $M$ = Co, Rh, Ir) are found to form two series of compounds [1], namely, for $n$ =1 and for $n$ = 2. The Ce-based compounds have been studied in great detail since they exhibit a variety of interesting properties such as heavy fermion behavior, superconductivity, antiferromagnetic ordering, pressure-induced superconductivity, etc [2]. Considerable amount of work has been reported for the $M$ = Co, $n$ =1 compounds. However, for $n$ =2, only studies on Ce compound have been reported [3]. We have attempted the synthesis of a series of compounds, $R_2CoIn_8$ ($R$ = rare earth). Here we report the magnetic and transport properties of these compounds with $R$ = Ce, Pr and Dy. The Ce compound shows heavy fermion behavior, in agreement with the reported study. The Pr compound does not show magnetic ordering, but exhibits a magnetic susceptibility at low temperatures reminiscent of crystalline electric field effects. The Dy compound shows magnetic field dependent transitions both in magnetization and specific heat measurements.

## 2. Experimental details

The $R_2CoIn_8$ compounds with $R$ = Ce, Pr, Dy were synthesized by arc melting under a high purity argon atmosphere on a water-cooled cooper hearth. The samples were wrapped in tantalum foils, sealed in evacuated quartz tube and annealed at 600$^{\circ}$C, for one month. All the samples were characterized by X-ray powder diffraction (Cu-K$_\alpha$ radiation) at room temperature. Magnetization measurements were performed using a SQUID magnetometer (MPMS, Quantum Design) or a Vibrating Sample Magnetometer (Oxford Instruments). Electrical resistivity was measured employing ac four-probe technique. Specific heat measurements were carried out by relaxation method in the temperature range of 2 K- 300 K in applied fields up to 9 T (PPMS, Quantum Design).

## 3. Results & discussion

X-ray diffraction patterns were analyzed using the Rietveld method. The diffraction pattern could be fitted on the basis of tetragonal Ho$_2$CoGa$_8$-type structure (space group $P4/mmm$). The lattice parameters obtained are $a$ = 4.640 Å, $c$ = 12.250 Å (for Ce$_2$CoIn$_8$), $a$ = 4.626 Å, $c$ = 12.206 Å (for Pr$_2$CoIn$_8$) and $a$ = 4.558 Å, $c$ = 12.013 Å

(for $Dy_2CoIn_8$), and are in good agreement with the reported values [4]. A small amount of unreacted In impurity was present in some of the compounds. Figure 1 shows the temperature dependence of electrical resistivity for $R_2CoIn_8$ compounds with $R$ = Ce, Pr and Dy. The resistivity behavior of $Ce_2CoIn_8$ is typical of a heavy Fermion compound. This is further confirmed by the specific heat measurements, the results of which are shown as an inset in Fig. 1. The estimated value of $\gamma$, from an extrapolation of the $C/T$ vs. $T^2$ plot (not shown) at low temperatures (5 to 20 K), was found to be 525 mJ/K$^2$mol-Ce, characteristic value of a heavy Fermion compound and in agreement with an earlier study [3]. A slope change is observed in the electrical resistivity of Dy compound in the vicinity of magnetic transition and the ordering temperature is confirmed by the magnetization and specific heat measurements, as shown in Figs. 2 and 3.

The low temperature part of the susceptibility for $Dy_2CoIn_8$ for various applied magnetic fields is shown in Fig. 2. The magnetization at low fields indicates the presence of two magnetic transitions, one transition at 17 K and the other near 5 K. The magnetic transition at 17 K appears to be of antiferromagnetic nature, and the transition near 5 K may be due to a spin reorientation. No apparent deviation was observed between the FC and ZFC magnetization measurements at 0.1 T (figure not shown), ruling out any possibility of spin canting. In order to trace out the origin of the second magnetization peak near 5 K, the magnetization was measured as a function of applied magnetic field, the plots of which are shown as the inset of Fig. 2. It is clear that two metamagnetic transitions occur at 3.6 T ($H_{M1}$) and 8.2 T ($H_{M2}$). This also supports our assumption that magnetic transition near 5 K may be due to the reorientation of the spins from the antiferromagnetically ordered state. The magnetization data as a function of temperature is obtained at various applied magnetic fields also show a similar feature. The transition broadens and the ordering temperature shifts from ~17 K at 0.1 T to ~15 K at 5 T (>$H_{M1}$) and the magnetic ordering near 5 K disappears altogether. Application of a field higher than $H_{M2}$ (9 T) shifts the ordering temperature further and saturation of magnetization is observed at low temperatures. This broadening in magnetization is expected since the system undergoes transition to spin re-oriented states with increased magnetization below the ordering temperatures at high fields, $H = 5$ T ($>H_{M1}$) and $H = 9$ T ($>H_{M2}$). The magnetic behavior of $Dy_2CoIn_8$ is also reflected in the specific heat data, the results of which are given in Fig. 3. In zero applied fields, the specific heat shows two peaks, in agreement with the low field magnetization measurements. As the field is increased, the position of the peak at 17 K gets shifted towards lower temperatures as expected for an antiferromagnetic transition whereas the peak near 5 K gets smeared out. The effect of metamagnetic transitions observed in this compound is evident from the decreased height of the specific heat peak at the ordering temperature.

Temperature variation of the inverse magnetic susceptibility of $Pr_2CoIn_8$ is given in Fig. 4. No magnetic transition is observed in this compound down to 2 K. However, its magnetic susceptibility deviates strongly from the Curie-Weiss behavior (shown as the straight line fit in Fig. 4) from around 80 K and then becomes nearly temperature independent below 30 K. The linear behavior of $M$ vs. $H$ plot at 2 K (inset of Fig. 4) excludes any possibility of ordered moments. Resistivity also shows a slope change at low temperatures, as shown in Fig. 1. Such a behavior is usually attributed to the crystalline electric field (CEF) effects [5], which remove the degeneracy of the ground state of the rare earth ion. The splitting of the (2$J$+1)-fold degenerate ground state of the $Pr^{3+}$ ion results in a singlet CEF ground state. At low temperatures the ground state is more populated than the excited states, giving rise to the temperature independent susceptibility. In the absence of magnetic ordering, this small effect becomes more apparent to be observed at low temperatures. We believe that the observed deviation of magnetic susceptibility from the Curie-Weiss behavior is due to the CEF effects. An estimation of the crystal field splitting in this compound is in progress.

In conclusion, we have presented some of the interesting properties exhibited by some members of the $R_2CoIn_8$ family of compounds. These properties include heavy fermion behavior ($R$ = Ce), possible CEF effects ($R$ = Pr), field-dependent magnetic ordering and metamagnetic transitions ($R$ = Dy).

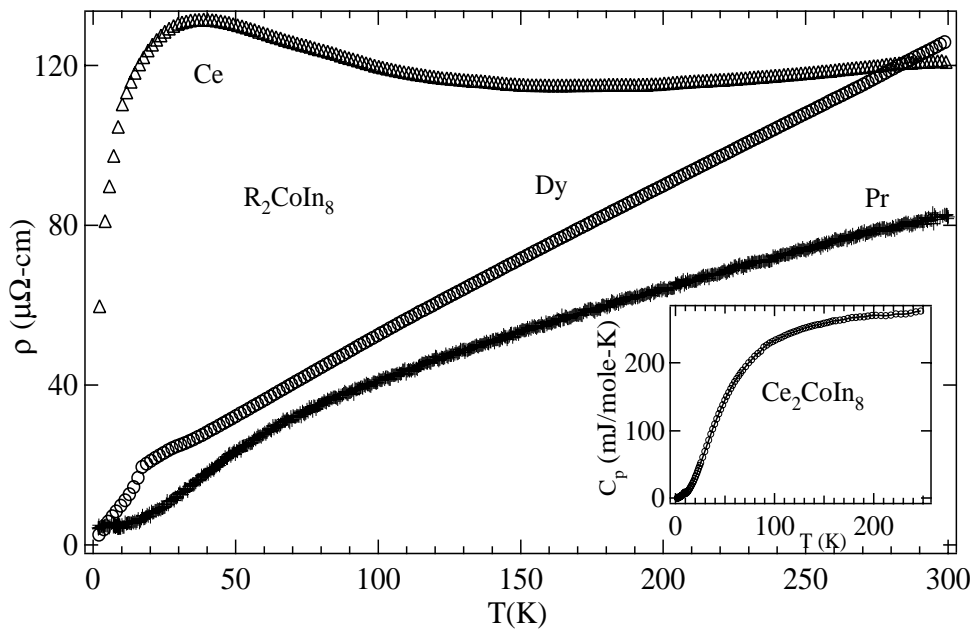

Fig. 1. Joshi et al

Fig.1.
Resistivity vs. temperature for $R_2CoIn_8$ ($R$ = Ce, Pr and Dy) compounds. The inset shows specific heat vs. temperature $Ce_2CoIn_8$.

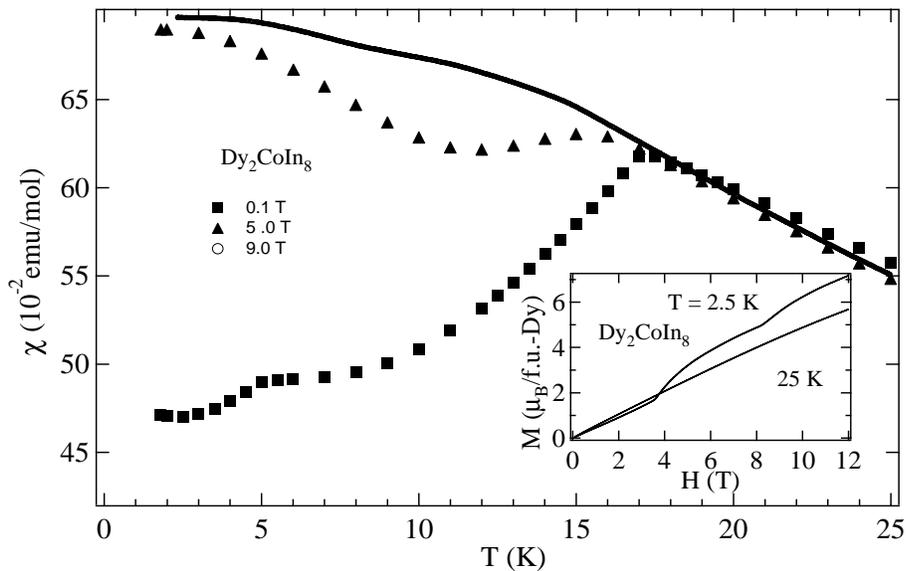

Fig. 2. Joshi et al

Fig. 2.
Magnetic susceptibility vs. temperature for $Dy_2CoIn_8$ at various applied magnetic fields. The inset shows $M$ vs. $H$ curves at 2.5 K and 25 K.

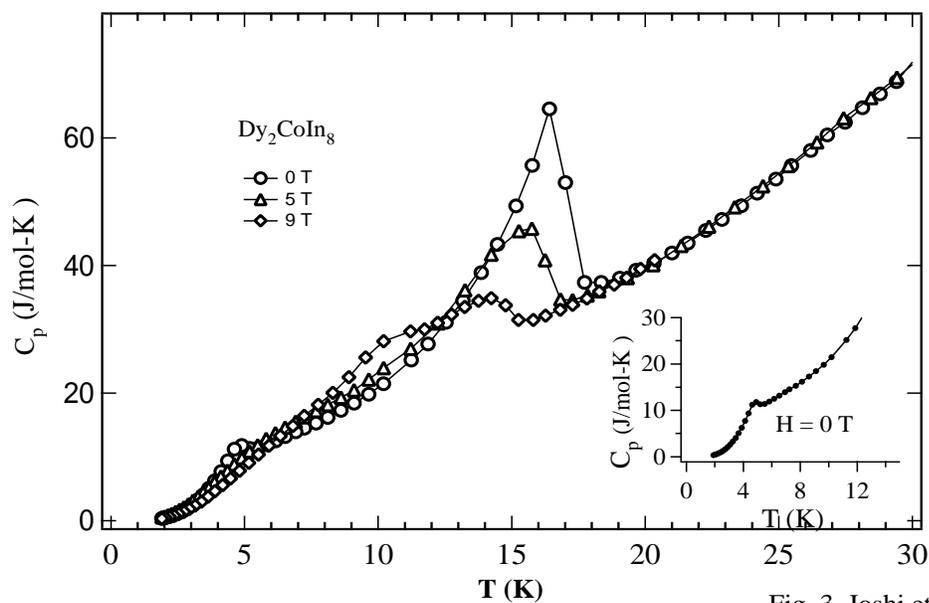

Fig. 3.
Heat capacity vs. temperature for Dy$_2$CoIn$_8$ at various applied magnetic fields. The inset shows the expanded version of the low temperature heat capacity indicating the magnetic transition at 5 K.

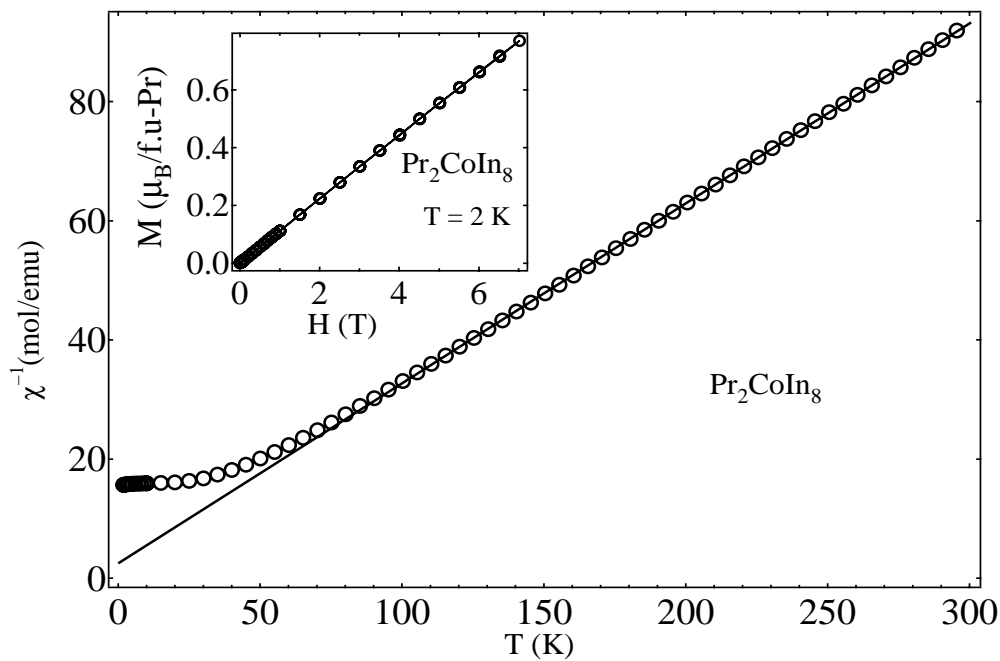

Fig. 4.
Temperature variation of the inverse magnetic susceptibility for Pr$_2$CoIn$_8$. The Curie-Weiss behavior is shown as the straight line fit. The inset shows the *M* vs *H* curve at 2 K